\documentclass[a4paper,fleqn,usenatbib]{mnras}

\usepackage[T1]{fontenc}
\usepackage{ae,aecompl}

\usepackage{graphicx}	
\usepackage{amsmath}	
\usepackage{amssymb}	
\usepackage{multicol}        
\usepackage{bm}         
\usepackage{pdflscape}  

\usepackage{txfonts}

\newcommand{\lapp}{\ifmmode\stackrel{<}{_{\sim}}\else$\stackrel{<}{_{\sim}}$\fi}
\newcommand{\gapp}{\ifmmode\stackrel{>}{_{\sim}}\else$\stackrel{>}{_{\sim}}$\fi}

\title{Determination of the Sun's Offset from the Galactic Plane Using Pulsars}

\author[]{
J. M. Yao,$^{1,3}$
R. N. Manchester,$^{2}$\thanks{E-mail:Dick.Manchester@csiro.au}
N. Wang$^{1,4}$
\\
$^{1}$Xinjiang Astronomical Observatory, Chinese Academy of
  Sciences, 150, Science 1-Street, Urumqi, Xinjiang 830011,
  China\\
$^{2}$CSIRO Astronomy and Space Science, Australia Telescope
  National Facility, P.O.~Box~76, Epping NSW~1710, Australia\\
$^{3}$University of Chinese Academy of Sciences, 19A Yuquan
  Road, Beijing 100049, China\\
$^{4}$Key Laboratory of Radio Astronomy, Chinese Academy of
  Science, Nanjing 210008, China
}

\date{Accepted XXX. Received YYY; in original form ZZZ}

\pubyear{2016}

\begin{document}
\label{firstpage}
\pagerange{\pageref{firstpage}--\pageref{lastpage}}
\maketitle

\begin{abstract}
We derive the Sun's offset from the local mean Galactic
plane($z_\odot$) using the observed $z$ distribution of young pulsars.
Pulsar distances are obtained from measurements of annual parallax, HI
absorption spectra or associations where available and otherwise from
the observed pulsar dispersion and a model for the distribution of
free electrons in the Galaxy. We fit the cumulative distribution
function for a ${\rm sech}^2(z)$ distribution function, representing
an isothermal self-gravitating disk, with uncertainties being
estimated using the bootstrap method. We take pulsars having
  characteristic age $\tau_c\lapp10^{6.5}$~yr and located within
4.5~kpc of the Sun, omitting those within the local spiral arm and
those significantly affected by the Galactic warp, and solve for
$z_\odot$ and the scale height, $H$, for different cutoffs in
$\tau_c$.  We compute these quantities using just the independently
determined distances, and these together with DM-based distances
separately using the YMW16 and NE2001 Galactic electron density
models. We find that an age cutoff at $10^{5.75}$~yr with YMW16
DM-distances gives the best results with a minimum uncertainty in
$z_\odot$ and an asymptotically stable value for $H$ showing
  that, at this age and below, the observed pulsar $z$-distribution is
  dominated by the dispersion in their birth locations. From this
sample of 115 pulsars, we obtain $z_\odot=13.4\pm$4.4~pc and
$H=56.9\pm$6.5~pc, similar to estimated scale heights for OB stars and
open clusters. Consistent results are obtained using the
independent-only distances and using the NE2001 model for the DM-based
distances.
\end{abstract}



\section{Introduction}\label{sec:intro}
It has long been known that the Sun is offset from the local mean
Galactic plane toward the North Galactic Pole. An early estimate of
the offset by \citet{tul42} based on an analysis of mainly local stars
gave $z_\odot=13.5\pm$1.7~pc. As listed in Table~\ref{tb:z_value},
over the last few decades a variety of astronomical objects and
methods have been used to estimate $z_\odot$.\footnote{See
  \citet{km17} for a more extensive list of historical $z_\odot$
  determinations.} In \citet{cv90}, the $z$ distribution of Wolf-Rayet
stars within 4.5~kpc of the Sun was fitted with a single,
self-gravitating and isothermal disk \citep{spi42,bah84}, giving
$z_\odot=15\pm$3~pc. The improved photometric accuracy and increased
sky coverage of the Palomar Sky Survey \citep{hl95} and the Sloan
Digital Sky Survey \citep{css+01} gave $z_\odot$ values a little
larger than previous results. \citet{mai01} used Hipparcos
trigonometric parallaxes to derive the $z$-distribution of local OB
stars, fitting it with an isothermal disk to obtain $z_\odot =
24.2\pm$1.8~pc. More recent studies have analysed the observed
distributions of OB stars, open clusters, Cepheid variables, HII
regions and giant molecular clouds. For example, \citet{jos07} used
three different methods in their analysis, obtaining values in the
range 6 -- 28~pc for OB stars with $d<$ 1.2~kpc and 13 -- 20~pc for
young open clusters with age $<10^{8.5}$~yr and $d<$ 4~kpc. An
asymmetric $z$ distribution of Cepheid variables with respect to the plane through the Sun was discovered by \citet{mtl09}. They obtained $z_\odot
= 26\pm$3~pc from a Gaussian fit to Cepheids with $d<$2~kpc.
\begin{table}
	\caption{Previous determinations of $z_\odot$
          based on different astronomical objects.}
	\label{tb:z_value}
	\begin{tabular}{ccc} 
		\hline
		Reference & Data sets &$z_\odot$ (pc)\\
		\hline
		\citet{tul42}&various stars&13.5$\pm$1.7\\
		\citet{cv90}&Wolf-Rayet stars&15$\pm$3\\
		\citet{hl95}&optical stars&20.5$\pm$3.5\\
		\citet{css+01}&optical stars&27$\pm$4\\
		\citet{mai01}&OB stars&24.2$\pm$1.8\\
		\citet{jos07}&OB stars&6-28\\
		             &open clusters&13-20\\
		\citet{mtl09}&Cepheid variables&26$\pm$3\\
                \citet{bf14} & open clusters & 18.5$\pm$1.2\\
		\citet{ok14} & magnetars& 13.9$\pm$2.5\\
                \citet{bb16b}&methanol masers&5.7$\pm$0.5\\
                     &HII regions&7.6$\pm$0.4\\
                     &giant molecular clouds&10.1$\pm$0.5\\
                \citet{jdpj16}&open clusters&6.2$\pm$1.1\\
		\hline
	\end{tabular}
\end{table}

As Table~\ref{tb:z_value} shows, recent analyses have tended to have
smaller uncertainties, although still with a significant spread of
values, largely because of the increased number of astronomical
objects, improved accuracy of the distances and consideration of
perturbing influences. \citet{bb16b} used 639 methanol masers, 878 HII
regions and 538 giant molecular clouds located in the inner region of
the Galaxy and fitted these data sets to a self-gravitating isothermal
disk. After considering the effects of the local spiral arm, the Gould
Belt and the Galactic warp, they obtained $z_\odot=5.7\pm$0.5~pc for
methanol masers, $z_\odot=7.6\pm$0.4~pc for HII regions and
$z_\odot=10.1\pm$0.5~pc for giant molecular clouds.  \citet{jdpj16}
used an almost complete sample of 1241 open clusters within 1.8~kpc of
the Sun and fitted an exponential profile to the data sets, resulting
in $z_\odot=6.2\pm$1.1~pc. An analysis by \citet{bf14} of a different
sample of open clusters gave a much larger value of $z_\odot$,
$18.5\pm1.2$~pc. Finally, \citet{ok14} analysed the $z$-distribution
of 19 magnetars by fitting to the cumulative distribution function
(CDF) and obtained $z_\odot=13.9\pm$2.5~pc.

Distances to astronomical objects are key for any analysis of the
Sun's offset from the Galactic plane. Pulsar distances can be
estimated from measurements of annual parallax, HI absorption spectra
and associations with globular clusters or supernova remnants and
optical observations of binary companion stars. At present, only 189
pulsars have such ``independent'' distances; these pulsars are listed
in the Appendix of \citet{ymw17} (Table 8 to Table 12). Fortunately,
for most pulsars we can estimate distances using interstellar
dispersion, quantified by the dispersion measure (DM):
\begin{equation}
    {\rm DM}=\int^D_0 n_e dl,
	\label{eq:DM_D}
\end{equation}
where $n_e$ is the free electron density and $D$ is the pulsar
distance, and a model for the Galactic $n_e$ distribution.  Until
recently, the NE2001 model
\citep{cl02}\footnote{\url{https://www.nrl.navy.mil/rsd/RORF/ne2001/}}
was the default choice for computing DM-based distances. However, we
have recently published the YMW16 model
\citep{ymw17}\footnote{\url{http://www.xao.ac.cn/ymw16/},\\
  \url{http://www.atnf.csiro.au/research/pulsar/ymw16/}}
which takes advantage of advances in our knowledge of Galactic
structure and a significant increase in the number of pulsars with
independent distance estimates over the past 15 years. In this work,
we consider three pulsar samples: a) pulsars with independent
distances, b) pulsars with independent distances plus DM-based
distances using the NE2001 model, and c) pulsars with independent
distances plus DM-based distances using the YMW16 model.
  
Since pulsars are relatively high-velocity objects
\citep[e.g.,][]{hllk05} and pulsar ages, based on the characteristic
age $\tau_c=P/(2\dot{P})$ where $\dot{P}$ is the period slow-down
rate, can be large, especially for millisecond pulsars, we need to
limit the age range of the pulsar sample used in the analyses. Only
for relatively young pulsars is the $z$ distribution dominated by
their birth location. For preliminary analyses, we adopt a limit of
$\tau_c <10^{6.5}$~yr, and then we investigate the effect of age by
fitting to samples with five different age limits in the range
$10^{5.25}$~yr and $10^{6.5}$~yr.

The arrangement of our paper is as follows. Data sets and analysis
methods are discussed in \S\ref{sec:methods}. In \S\ref{sec:influence}
we consider the effect of the Gould Belt, the Local Arm and the
Galactic warp on our analysis. In \S\ref{sec:result} we consider the
effect of the different $\tau_c$ cutoffs and present our final results
and we summarise our conclusions in \S\ref{sec:conclusion}.

\section{Methods}\label{sec:methods}
\begin{figure}
	\includegraphics[width=6.0 cm, angle=270]{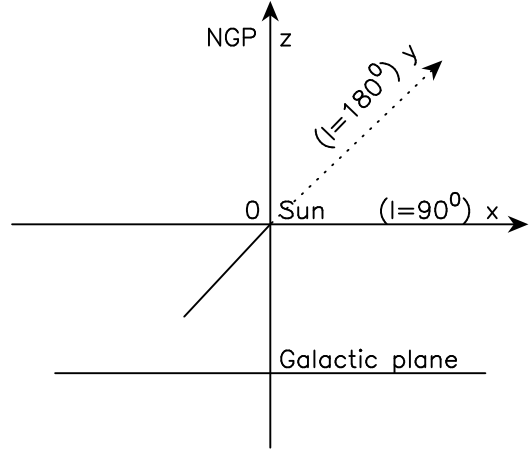}
    \caption{Coordinate system for analysis of the pulsar $z$ distribution.}
    \label{fig:Coordinate}
\end{figure}

For this analysis, we adopt a coordinate system with origin at the Sun,
with the $x$ axis toward $l=90\degr$, $y$ axis toward $l=180\degr$ and
$z$ axis toward the North Galactic Pole
(Figure~\ref{fig:Coordinate}). Note that this coordinate system is not
the same as the Galactocentric coordinate system in which $z_\odot$ is
defined, where origin is at the Galactic Centre and the $X-Y$ plane is
the Galactic plane. In the heliocentric system, coordinates of pulsars
(or other objects) are ($x$, $y$, $z$) = ($d\sin l \cos b$, $d\cos l
\cos b$, $d\sin b$), where $d$ is distance from the Sun and $l$, $b$
are the Galactic longitude and latitude, respectively. In the
Galactocentric coordinate system, the Sun is located at
$R_\odot=$8300~pc \citep{brm+11}. 

A variety of different methods have been employed to estimate
$z_\odot$ as discussed in \S\ref{sec:intro}. If the distances to
astronomical objects are available, fitting a function to the observed
$z$ distribution is the most common method \citep[e.g.,][]{cv90,
  bb16b, jdpj16}. Many of these previous works adopted a
self-gravitating isothermal disk model in which the $z$ distribution
is described by ${\rm sech}^2(z)$. The Galactic electron density
models, NE2001 and YMW16, also adopt this function for $z$
distributions. Consequently, we also use it and model the number density
distribution of pulsars in $z$ by the following equation:
\begin{equation}
N(z)=N_0\;{\rm sech}^2\left(\frac{z+z_\odot}{H}\right)
\label{eq:distrib}
\end{equation}
where $N_0$ is number density in the Galactic plane, $z_\odot$ is the
distance between the Sun and the Galactic plane, and $H$ is
the scale height of pulsars with respect to the Galactic plane.

As discussed by \citet{ok14}, fitting Equation~\ref{eq:distrib}
  directly to binned data for the $z$-distribution fails when the
  sample is small in number. In such cases, it is better to fit to the
  CDF which, for the ${\rm sech}^2(z)$ distribution given by
  Equation~\ref{eq:distrib}, is
\begin{equation}
F(z;z_\odot,H)=\frac{1}{1+\exp{-\frac{2(z+z_\odot)}{H}}}
\label{eq:cdf}
\end{equation}
where F is in the range [0,1] for $z$ in the range $\pm \infty$. We
use the Levenberg-Marquardt algorithm \citep{ptvf92} to fit for
$z_\odot$ and $H$. Because of the small samples, the uncertainties in
these parameters are dominated by sample variance rather than the
formal fit uncertainties. Consequently, we use the bootstrap method
\citep{efr81} to calculate parameter uncertainties. Our bootstrap
procedure for a sample of N $z$-values is as follows:
\begin{enumerate}
\item Randomly select a $z$ value from the sample N times to build up a new sample of N values
\item Solve for $z_\odot$ and $H$ using this sample and store the results
\item Repeat steps i-ii 500 times, generating 500 $z_\odot$ and $H$ values
\item Take the rms deviation of these values about the mean as the
  uncertainty in the parameters derived from a fit to the original
  sample.
\end{enumerate} 

We have applied these methods to several different samples of the
known pulsar population, selected as follows.  We firstly take the
sample of 189 pulsars with independent distances given by
\citet{ymw17} and select pulsars with known age ($\tau_c$) and DM, and
omit pulsars associated with globular clusters, giving a total of 161
pulsars that we can use for modelling the $z$ distribution. Especially
after limiting the age range, this sample is very small, and so we
need to supplement it using pulsars with distances obtained from the
DM and and a Galactic $n_e$ model.  From the ATNF Pulsar
Catalogue\citep[V1.54,][]{mhth05},
\footnote{\url{http://www.atnf.csiro.au/research/pulsar/psrcat}},
and applying the same selection criteria, we obtain a list of 1923
pulsars. For the 1762 of these without independent distances, we can
use either the NE2001 or YMW16 Galactic $n_e$ models to estimate
distances and hence $z$ values from the observed DM.

Since we are interested in the offset of the Sun from the local mean
Galactic plane, we further restrict our samples to pulsars having
estimated distances less than 4.5~kpc. This leaves 134 pulsars in the
sample with independent distances, 936 from the sample with NE2001
distances and 911 from the sample with YMW16 distances. 
  
\begin{figure}
	\includegraphics[width=6.6 cm, angle=270]{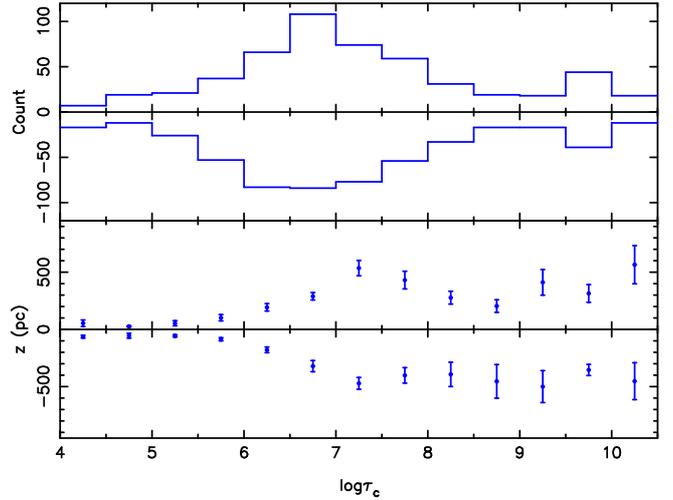}
    \caption{The upper panel shows the histogram of number of pulsars
      vs log$\tau_c$ separately for pulsars with $z>0$ and $z<0$ with
      model distances based on YMW16. The lower panel shows the mean
      $z$ for each bin sub-sample. Only pulsars within 4.5~kpc of the
      Sun are included in the sample.}
    \label{fig:age_z}
\end{figure}

As discussed in \S\ref{sec:intro} we need to limit the age range of
pulsars included in the analysis. Figure~\ref{fig:age_z} shows the
distribution of pulsars (using the local YMW16 sample described above)
and their mean $z$ separately for $z>0$ and $z<0$ as a function of
characteristic age. The upper panel of Figure~\ref{fig:age_z} shows
that pulsars cluster in two groups, one correponding to normal pulsars
with $\log\tau_c \lapp 8.5$ and the other at larger $\tau_c$
corresponding to millisecond pulsars. From the bottom panel of
Figure~\ref{fig:age_z}, the mean $z$ increases with increasing
$\log\tau_c$ up to $\log\tau_c < 7.0$ showing the effect of pulsar
velocities. We select pulsars from the samples described above with
$\log\tau_c \leq 6.5$ as base samples for further investigations of
the effect of pulsar age and other factors on the derived $z_\odot$
and $H$. These contain 63, 289 and 341 pulsars for the independent,
NE2001 and YMW16 distances, respectively. 

\section{Factors influencing the $z$ distribution}\label{sec:influence}
\subsection{Gould Belt and Local arm}
It has long been recognised that the Sun is located interior to a ring
of relatively young stars and stellar clusters known as the Gould Belt
\citep[e.g.,][]{eca06}. The Gould Belt has a radius of about 300~pc
and is centred about 100~pc from the Sun toward the Galactic
anti-centre with an inclination to the plane of about $18\degr$
\citep[e.g.,][]{pc03}. Although \citet{jdpj16} omitted 26 open
clusters believed to be associated with the Gould Belt from their
analysis, we see no effect of the Gould Belt on the observed pulsar
distribution, most probably because only a few pulsars are located
within the Gould Belt region. 

\begin{figure}
\includegraphics[width=7.0 cm, angle=270]{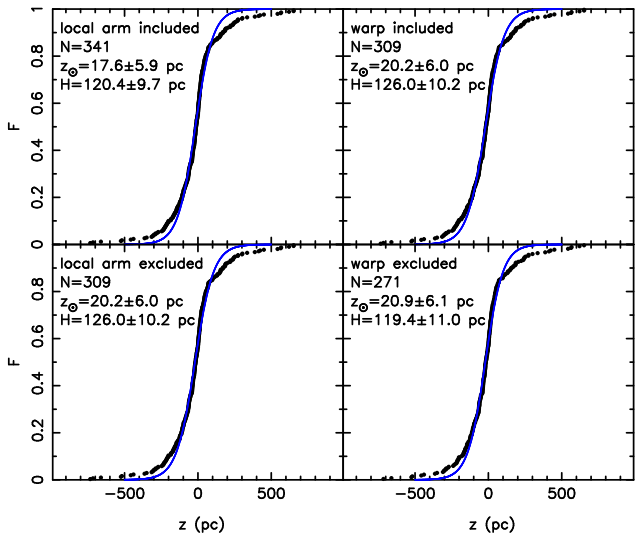}
\caption{Cumulative $z$-distributions for four pulsar
  samples as described in the text. The effect of the Local Arm is
  shown in the left panels and the right panels show the effect of the
  Galactic warp. In each case, the black dots show the observed distribution and the blue line shows the fit of Equation~\ref{eq:cdf}.}
    \label{fg:arm_warp}
\end{figure}

Studies of the Galactic distribution of young objects \citep[see,
  e.g.,][]{hh14} show evidence of a weak spiral feature close to the
Sun, known as the Local Arm. From an analysis of 29 methanol masers
believed to be associated with the Local Arm, \citet{bb16b} found that
the Local Arm is centred $25-35$~pc above the Galactic plane.
To investigate the effect of the Local Arm on our results, we
  adopt the age-limited sample with independent and YMW16-based
  distances containing 341 pulsars described above and fit the
  CDF (Equation~\ref{eq:cdf}) to the
  observed pulsar cumulative $z$ distribution. The results of the fits
  are presented in the top left panel of Figure~\ref{fg:arm_warp}.

Using the definitions of \citet{ymw17}, a total of 32 pulsars from
this sample have perpendicular distances from the Local Arm $s_{\rm
  a}<300$~pc. If we exclude these pulsars and repeat the fit to the
cumulative distribution, we obtain the results presented in the bottom
left panel of Figure~\ref{fg:arm_warp}. The larger value of $z_\odot$
for the sample omitting Local Arm pulsars (20.2$\pm$6.0~pc versus
17.6$\pm$5.9~pc) is consistent with the Local Arm being located above
the mean Galactic Plane as found by \citet{bb16b}. Although the
effect of the Local Arm is marginal, to avoid any bias we choose to
omit the Local Arm pulsars from subsequent analyses.

\subsection{Galactic Warp}
It is well known that the outer Galactic disk has a substantial warp
\citep[e.g.,][]{ufm+14}. Following \citet{rrdp03} and using a
Galactocentric ($X$, $Y$, $Z$) coordinate system, \citet{ymw17} have modeled the $Z$ offset due to the Galactic warp as follows:
\begin{equation}\label{eq:warp}
  Z_w = Z_c \cos(\phi - \phi_w),
\end{equation}
\begin{equation}
  Z_c = \gamma_w(R - R_w),
\end{equation}
 where $Z_c$ and $\phi_w$ respectively represent the maximum
  warp offset at a given Galactocentric radius $R>R_w$ and the
  direction for which the warp offset is maximum, and $\phi$ is
measured counterclockwise from the $+X$ direction, parallel to
$l=90\degr$. For $\phi_w=0$ and $R_w=8400$~pc, \citet{ymw17} obtain
$\gamma_w=0.14\pm0.066$ and we adopt these parameters here. The warp
affects the disk at $R>R_w$ and is toward $+Z$ in the $+X$
direction and toward $-Z$ in $-X$ direction. In
Figure~\ref{fg:arm_warp}, the top right panel is a repeat of the
bottom left panel which ignores the warp, whereas for the bottom right
panel, we omit the 38 pulsars for which $Z_w>$10~pc. Again, although
the effects of omitting the warp-affected pulsars are marginal, we
choose to exclude them from the samples used for further analysis. For
the independent distances, independent plus NE2001 model distances and
independent plus YMW16 model distances, the resulting sample sizes are
45, 222 and 271 pulsars, respectively.

\section{Results and Discussion}\label{sec:result}
\begin{table*}
	\caption{Values of $z_\odot$ and $H$ from fits to cumulative
          $z$-distributions with samples based on just independent
          distances, independent distances plus NE2001 model
          distances, and independent distances plus YMW16 model
          distances with different cut-off ages. Uncertainties are
          estimated using a bootstrap method.}
	\label{tb:age_result}
	\begin{tabular}{cccccccccccc}
	\hline
		& & Independent & & & &NE2001& & & &YMW16& \\
	   \cline{2-4}\cline{6-8}\cline{10-12}
		Samples & N &$z_\odot$& $H$ & &N &$z_\odot$& $H$ & &N &$z_\odot$& $H$ \\
                  &      & pc          & pc & & &pc& pc & & &pc&pc\\
		\hline
		 log$\tau_c<$ 6.5  & 45&30.2$\pm$15.5 &111.5$\pm$29.3 & & 222&20.0$\pm$9.7&181.9$\pm$14.3& &271&20.9$\pm$6.1&119.4$\pm$11.0 \\
		 log$\tau_c<$ 6.0  & 28&20.3$\pm$13.3 &72.5$\pm$25.4  & & 113&17.1$\pm$8.3&99.5$\pm$12.3&  &151&17.1$\pm$5.2&74.9$\pm$8.1   \\
		 log$\tau_c<$ 5.75 &21 & 7.1$\pm$12.1&44.6$\pm$26.5   & & 78&12.0$\pm$6.6&68.2$\pm$12.8&   &115&13.4$\pm$4.4&56.9$\pm$6.5   \\
		 log$\tau_c<$ 5.5  &15 &16.2$\pm$18.2&57.0$\pm$35.9   & & 53&13.2$\pm$8.0&54.8$\pm$14.1&   &79&15.9$\pm$6.0&54.5$\pm$7.0    \\
		 log$\tau_c<$ 5.25 &14 & 21.5$\pm$30.2&67.6$\pm$74.7  & & 44&11.5$\pm$9.3&58.5$\pm$15.9&   &66&12.5$\pm$6.4&53.9$\pm$6.4    \\
		\hline
	\end{tabular}
\end{table*}

\begin{figure}
\includegraphics[width=12.0 cm, angle=270]{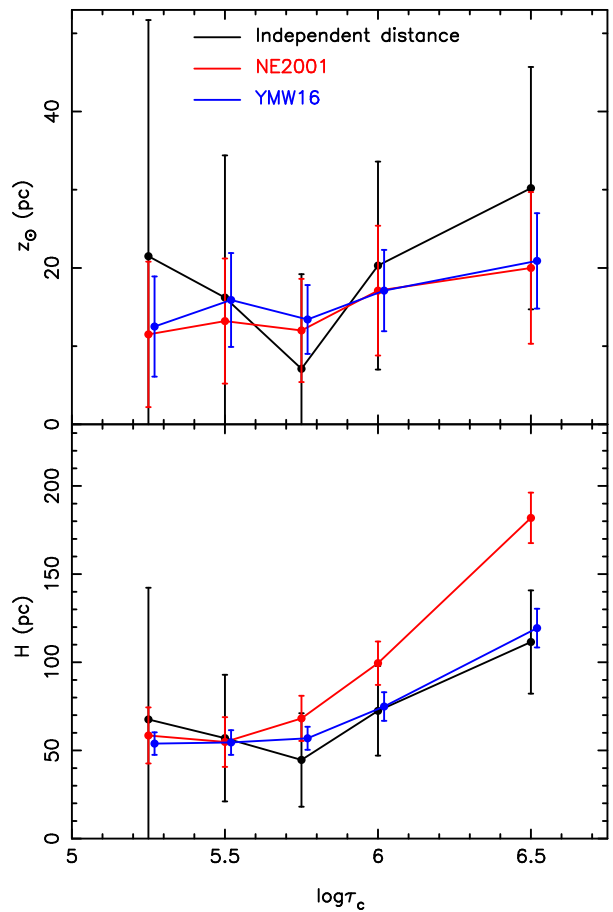}
\caption{Results of fitting for $z_\odot$ and $H$ plotted against age
  cutoff. Results from the sample with just independent distances are
  shown in black, those with NE2001 model distances are red and those
  with YMW16 model distances are blue. For clarity, the YMW16 points
  are offset by 0.02 in $x$. Uncertainties are estimated using a
  bootstrap method.}
    \label{fg:Age_H_z}
\end{figure}

Taking the final samples discussed in the previous section, we fit the
cumulative $z$-distributions for $z_\odot$ and $H$ using the CDF given
by Equation~\ref{eq:cdf} for cutoffs in $\log\tau_c$ between 5.25 and
6.5. The results are tabulated in Table~\ref{tb:age_result} and
plotted in Figure~\ref{fg:Age_H_z}. Uncertainties are estimated by
using the bootstrap method described in
Section~\ref{sec:methods}. These are generally a factor of three or so
larger than the uncertainties given by the least-squares fit, but we
believe that they better represent the true uncertainty in the derived
values, especially for the smaller samples. Because of the much
smaller sample sizes, results for the independent-only samples are
much more uncertain, but never-the-less consistent with the larger
samples which include the DM-based distances.

With decreasing age cutoff, the scale height $H$ stabilises at a
little less than 60~pc for cutoffs (in $\log\tau_c$) of 5.75 and
lower. This shows that, for pulsars younger than about 550 kyr, the
observed $z$-distribution is dominated by the dispersion in birth
location (in $z$) rather than the dispersion resulting from pulsar
space velocities over their lifetime.  It is notable that the derived
scale height increases much more rapidly with increasing age for the
NE2001 sample compared to both the independent and YMW16
samples. This is most probably a consequence of the typically smaller
  spiral-arm densities in the NE2001 model compared to the
  YMW16 model, especially for the Carina arm. 

An age cutoff of $10^{5.75}$~yr also gives the smallest uncertainties in the
derived $z_\odot$ for all samples. These minima result from the
competing effects of tighter distributions and smaller sample sizes as
the age cutoff is reduced.

\begin{figure}
\includegraphics[width=12.0 cm, angle=270]{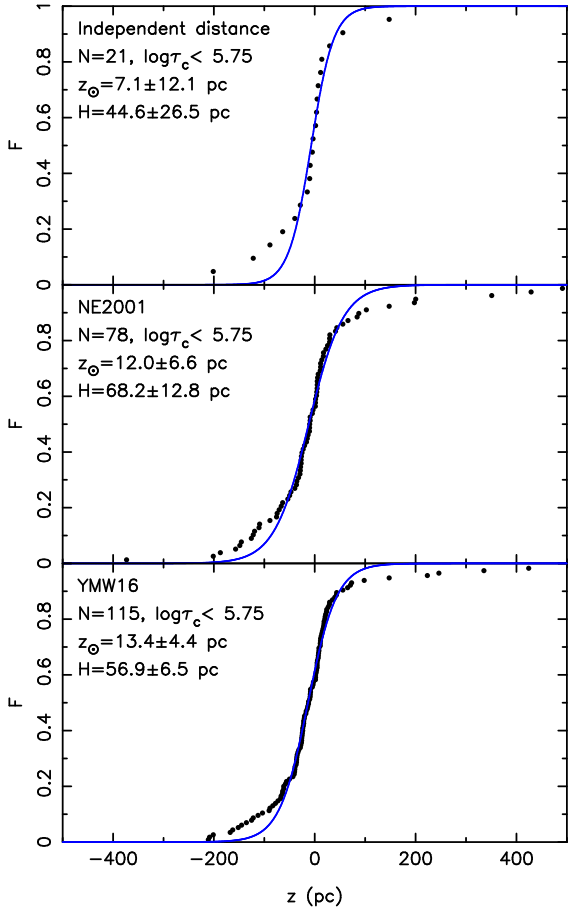}
\caption{Cumulative $z$-distributions for the final samples with only
  independent distances (top), NE2001 distances (middle) and YMW16
  distances (bottom). In each case, the black dots show the observed distribution
  and the blue line is the best-fit of Equation~\ref{eq:cdf}.}\label{fg:best_cdf}
\end{figure}

Figure~\ref{fg:best_cdf} shows the observed and fitted CDFs for the
three samples with age cutoff at $10^{5.75}$~yr. All of the observed
CDFs have a similar form although the effects of the different sample
sizes are evident. While the bulk of the population is well modelled
by the self-gravitating isothermal disk, with most pulsars lying
within 100~pc of the Galactic plane, there is a significant number of
outliers at larger $z$ for all three samples. The fact that these
outliers are present in the sample of independent distances shows that
they do not result from deficiencies in the $n_e$ models, but
represent a distinct high-$z$ population. Whether this results from a
population of high-$z$ birth locations or from a high-velocity tail on
the pulsar velocity distribution is unclear. In any case, it has
little influence on the fitted values of $z_\odot$ and $H$.

Of the three samples with age cutoff at $10^{5.75}$~yr, the one using
YMW16 model distances gives the smallest uncertainties. We therefore
adopt the results from fitting to the sample with YMW16-based
distances and age cutoff at $10^{5.75}$~yr as our final results, viz.,
as defined by Equation~\ref{eq:distrib}, the offset of the Sun from
the local mean Galactic plane, $z_\odot=13.4\pm4.4$~pc, and the scale
height in $z$ of local pulsars with $\tau_c\le 10^{5.75}$~yr,
$H=56.9\pm6.5$~pc.

\begin{figure}
\includegraphics[width=8.5 cm, angle=270]{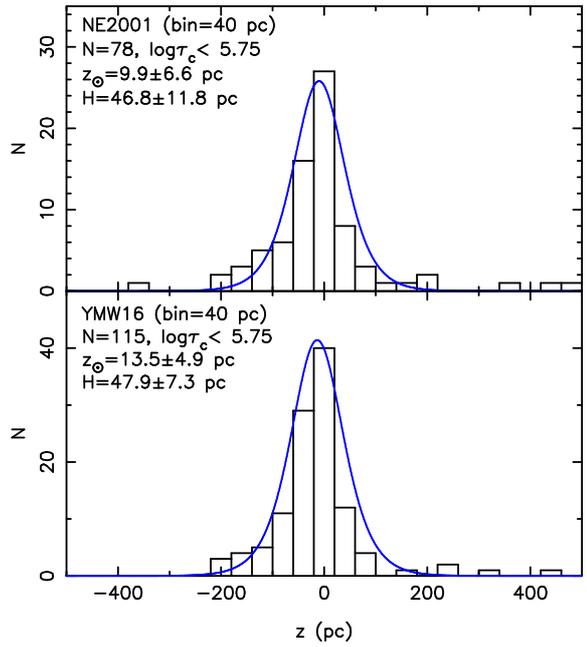}
\caption{Histograms of the $z$-distributions for the final samples
  with NE2001 distances (upper), YMW16 distances (lower). The blue
  line is the best-fit of Equation~\ref{eq:distrib} for each
  case.}\label{fg:best_bin}
\end{figure}

As a consistency check, we have directly fitted the sech$^2$
distribution function (Equation~\ref{eq:distrib}) to binned histograms
of the observed $z$ distributions for the final YMW16 and NE2001
samples; there are insufficient pulsars in the independent-distance
sample for this method to work reliably. The results are shown in
Figure~\ref{fg:best_bin}. The derived values of $z_\odot$ are very
similar to those from fitting of the cumulative histograms
(Table~\ref{tb:age_result} and Figure~\ref{fg:best_cdf}). Estimates of
pulsar scale height $H$ tend to be smaller with the histogram fits, but
still quite consistent within the uncertainties with those from the
CDF fits. Evidently, outliers have less effect on the histogram fits.

We have also investigated the effect of uncertainties in the adopted
distances for pulsars in our sample. Uncertainties in the independent
distances are tabulated in \citet{ymw17}. Uncertainties in DM-based
distances are harder to assess, but the analysis in \citet{ymw17}
(Table 5) suggests an rms deviataion of about 45\% for YMW16 model
distances and 60\% for NE2001 model distances. We used a Monte Carlo
procedure to investigate the effect of these uncertainties on the
results, randomly choosing a distance within the uncertainty
  range for each pulsar in the $\log\tau_c<$5.75 samples, fitting for
  $z_\odot$ and $H$ from the cumulative distributions, and repeating
  this 500 times. The rms deviations of the resulting $z_\odot$
  distributions were just 1.1~pc and 0.75~pc for the NE2001 distances
  and YMW16 distances respectively, much less than the quoted rms
  uncertainties (Table~\ref{tb:age_result}) which are dominated by
  sample variance and measured using the bootstrap method. To a lesser
  extent, the same is true for the distributions in $H$.

Our final result for the solar offset, $z_\odot=13.4\pm$4.4~pc, is
well within the range of other recent determinations of this parameter
(Table~\ref{tb:z_value}) and essentially identical to the result of
\citet{ok14} from a fit to the $z$ distribution of magnetars. In a
recent paper, \citet{km17} compiled 55 estimates of $z_\odot$ made
over the past century. The median of these, 17$\pm$2~pc, is consistent
with our result. These agreements show that pulsars, despite their
relatively large space velocities, can give a reliable independent
measure of the Sun's offset from the local mean Galactic plane.

Similarly, our final result for the scale height of local
pulsars with $\tau_c\le 10^{5.75}$~yr, $56.9\pm$6.5~pc, is
  consistent with the derived scale height of $60\pm$2~pc for open
  clusters \citep{jdpj16} and $61.4^{+2.7}_{-2.4}$~pc for OB stars
  \citep{jos07}. This again shows that, for these relatively young
  pulsars, the space velocity of pulsars does not significantly affect
  their $z$ distribution. Scale heights for methanol masers, HII
  regions and giant molecular clouds determined by \citet{bb16b} are
  much smaller, in the range $34-40$~pc.\footnote{\citet{bb16b} use a
    scale factor of $\sqrt{2}H$ for an isothermal self-gravitating
    disk whereas we simply use $H$. We have therefore multiplied their
    quoted scale heights by $\sqrt{2}$ for comparison with our
    results.} This reflects the younger age of these Galactic
  components and the fact that pulsars are born at a later stage of
  evolution.

It is important to note that the YMW16 Galactic $n_e$ model includes
an offset of the Sun from the Galactic plane of +6.0~pc, whereas the
NE2001 model has no such offset. The agreement of the derived values
of $z_\odot$ using the two models for estimation of pulsar distances
where no independent distances are available, illustrated in
Figure~\ref{fg:Age_H_z}, shows that the inclusion of a non-zero
$z_\odot$ in YMW16 has no effect on the results derived in this
analysis. Other factors dominate the variations in pulsar distances
derived from the pulsar DM and Galactic $n_e$ models.

\section{Conclusions}\label{sec:conclusion}
Using independent distances for pulsars and, where these are not
available, distance estimates based on the pulsar dispersion measures
and two models for the Galactic $n_e$ distribution (NE2001 and YMW16)
we have fitted the observed $z$ distribution of young pulsars to
estimate the offset of the Sun $z_\odot$ from the local mean Galactic
plane.  After limiting the distance of pulsars from the Sun
($d<4.5$~kpc), omitting pulsars located within the local spiral arm
and those which may be affected by the Galactic warp ($Z_w>10$~pc),
and taking only young pulsars ($\tau_c<10^{5.75}$~yr), we derive
$z_\odot=13.4\pm$4.4~pc. This result is consistent with recent
determinations of $z_\odot$ using other young tracers of the Galactic
disk such as OB stars, open clusters, methanol masers, HII regions and
giant molecular clouds. It is also independent of which Galactic $n_e$
model is used to estimate pulsar distances from their dispersion
measure. The derived scale height for pulsars with $\tau_c\le
  10^{5.75}$~yr, $56.9\pm$6.5~pc, is dominated by the pulsar birth
  locations and is comparable to the observed scale height of OB
stars and open clusters, but about 50\% larger than that of HII
regions, methanol masers and giant molecular clouds.

\section*{Acknowledgements}
We thank the referee for helpful comments that have resulted in
significant improvements to the paper and members of the Pulsar Group
at XAO for useful discussions . This work was supported by National
Basic Research Program of China (973 Program 2015CB857100), the
Strategic Priority Research Programme (B) of the Chinese Academy of
Science (No. XDB09000000) and the National Natural Science Foundation
of China (No. 11373006).



\bibliographystyle{mnras}





\bsp	
\label{lastpage}
\end{document}